\documentclass[runningheads,a4paper]{llncs}
\usepackage{amsmath,amssymb}
\usepackage{epsfig}
\usepackage{listings}
\usepackage{multirow}
\usepackage{array}
\usepackage{xcolor,colortbl}

\definecolor{Gray}{gray}{0.85}
\newcolumntype{C}[1]{>{\centering\let\newline\\\arraybackslash\hspace{0pt}}m{#1}}

\begin{document}

\mainmatter

\title{DISA at ImageCLEF 2014 Revised: Search-based Image Annotation with DeCAF Features}
\titlerunning{Search-based Image Annotation with DeCAF Features}

\author{%
Petra Budikova \and
Jan Botorek \and
Michal Batko \and
Pavel Zezula}

\institute{Masaryk University, Brno, Czech Republic}

\maketitle

\begin{abstract}

This paper constitutes an extension to the report on DISA-MU team participation in the
ImageCLEF 2014 Scalable Concept Image Annotation Task as published in~\cite{BudikovaCLEF14}.
Specifically, we introduce
a new similarity search component that was implemented into the system, report on
the results achieved by utilizing this component, and analyze the influence of
different similarity search parameters on the annotation quality.
\end{abstract}

\section{Introduction}

With the continuous growth of popularity of multimedia data, it is nowadays obvious that
effective tools for multimedia storing, indexing and retrieval are much needed.
To encourage development of such tools for the image data domain, the ImageCLEF
Lab offers every year several challenges that reflect current open issues
in the image processing field. In 2014, one of these challenges was the Scalable Concept Image Annotation
task.

A team from DISA Laboratory\footnote{http://disa.fi.muni.cz} at Masaryk University was one
of the participants who submitted a solution for the 2014 Scalable Concept Image Annotation
task and whose results were evaluated within the contest. However, shortly after the evaluation we adopted
a new measure of visual similarity between images and replaced the respective similarity
search module of our annotation system. This led to significant improvements
of annotation quality in all evaluation measures, which we believe could be of interest
for the research community. We report the new results in this paper and furthermore
analyze the influence of different settings of content-based retrieval on the performance
of the annotation system.

\section{ImageCLEF Scalable Concept Image Annotation Task}

The problem offered by 2014 Scalable Concept Image Annotation (SCIA) challenge~\cite{ImageCLEF2014Overview,CLEF14AnnotationOverview}
is a standard annotation task, where relevant concepts from a fixed set of candidate concepts need to
be assigned to an input image. The {\em input images}
are not accompanied by any descriptive metadata such as EXIF or GPS, so that
only the visual image content can serve as annotation input. For each
test image, there is a {\em list of SCIA concepts} from which the relevant ones
need to be selected. Each concept is defined by one keyword, a link
to relevant WordNet nodes, and, in most cases, a link to a relevant Wikipedia page.

As the 2014 SCIA challenge focused especially on the concept-wise scalability of
annotation techniques, the participants were not provided with hand-labeled training data
and were not allowed to use resources that require significant manual preprocessing.
Instead, they were encouraged to exploit data that can be crawled from the web or otherwise easily obtained,
so that the proposed solutions should be able to adapt easily when the list of
concepts is changed.
Accordingly, the training dataset provided by organizers consisted of 500K images downloaded from
the web, and the accompanying web pages.
The raw images and web pages were further preprocessed by competition organizers to ease the participation in the task,
resulting in several visual and text descriptors as detailed in~\cite{CLEF14AnnotationOverview}.

The actual competition task consisted of annotating 7291 images with different concept lists.
Altogether, there were 207 concepts, with the size of individual concept lists ranging from
40 to 207 concepts. Prior to releasing the test image set, which became available a month before the
competition deadline, participants were provided with a development set of query images and concept lists, for which a ground truth of relevant
concepts was also published. The development set contained 1940 images and only 107 concepts
out of the final 207.

\section{DISA at ImageCLEF 2014: The Search-based Solution for Scalable Image Annotation}
\label{sec:related-work}

The DISA team participated in the SCIA task with a solution based on the MUFIN Image Annotation software, a tool for general-purpose image
annotation~\cite{BatkoBBZ13}. The
MUFIN Image Annotation tool follows the search-based approach to image annotation, exploiting content-based
retrieval in a large image collection and a subsequent
analysis of descriptions of similar images.

\begin{figure}[t]
\begin{center}
\includegraphics[width=\textwidth]{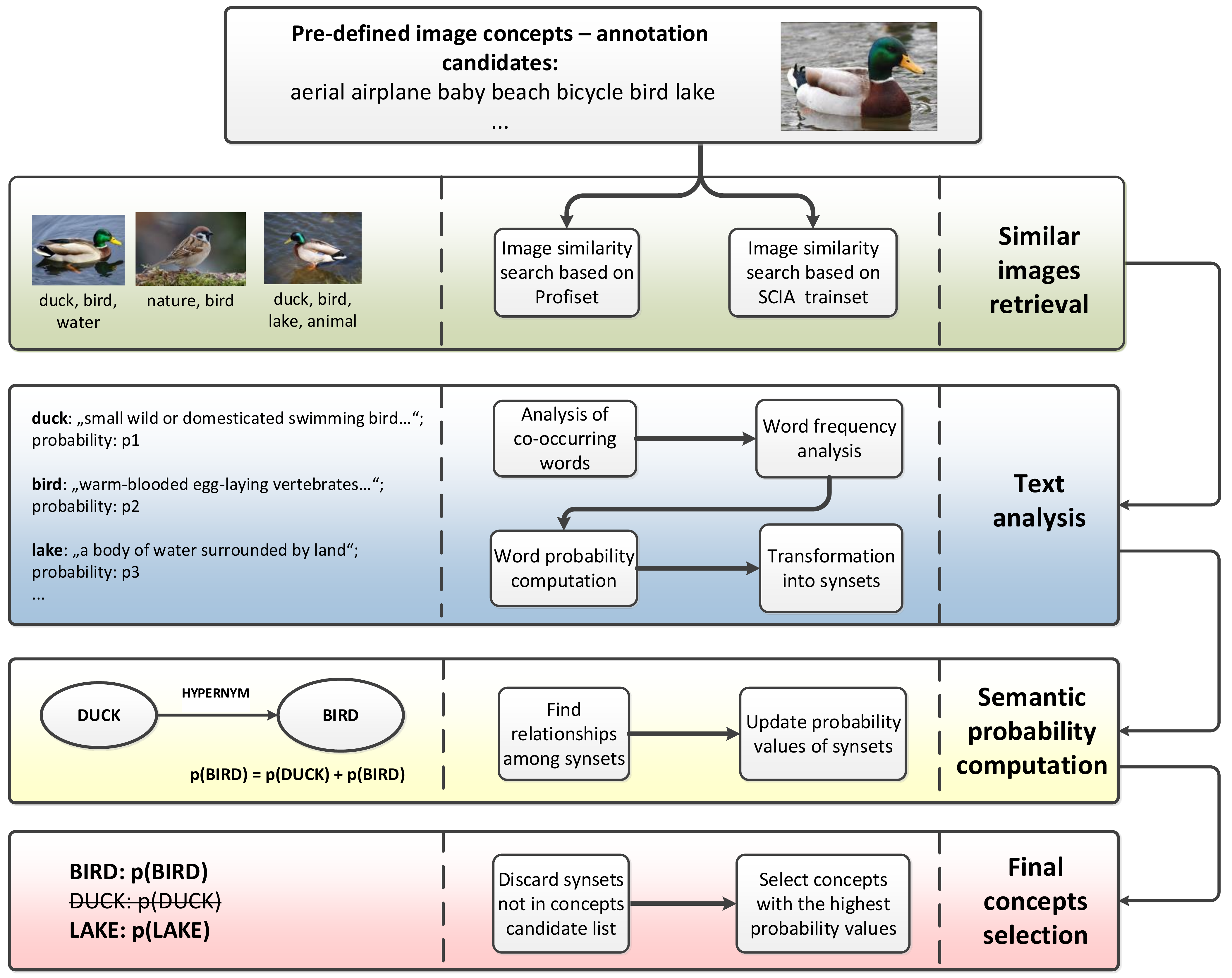}
\end{center}
\caption{\label{fig:Overview}Architecture of the DISA solution for ImageCLEF 2014}
\end{figure}

The general overview of the solution developed for the SCIA task is provided in Figure~\ref{fig:Overview}.
In the first phase, the annotation tool retrieves visually similar images from a suitable image collection.
Next, textual descriptions of similar images are analyzed with the help of various semantic resources. The text is split into separate words and
transformed into WordNet synsets, which are expanded and enhanced by semantic relations. The probability of relevance of each synset is computed with respect
to the initial probability value assigned to that synset and the types and amount of relations formed with other synsets.
Finally, synsets linked to the candidate concept words (i.e. the words in the list of concepts provided with the particular test image) are
ordered by probability and a fixed number of top-ranking ones is selected as the final image description.

In the following sections, we briefly outline the two main components of the annotation system,
focusing on details salient for our further discussion of improvements introduced after ImageCLEF 2014
competition deadline.
A more detailed description of the whole DISA annotation system can be found in~\cite{BudikovaCLEF14}.

\subsection{Phase 1: Retrieval of similar images}

The search-based approach to image annotation is based on the assumption that
in a sufficiently large collection, images with similar content to any given query
image are likely to appear. If these can be identified by a suitable content-based
retrieval technique, their metadata such as accompanying texts, labels, etc. can
be exploited to obtain text information about the query image.

\subsubsection{Image Collections}
In our solution for the SCIA challenge, we utilized two annotated image collections.
The Profiset collection~\cite{Budik11} contains 20M high-quality images with rich keyword annotations obtained from a photo-stock website,
which are freely available for research purposes. The data contained in the Profiset collection
was created manually, however this labor was not focused on providing training data for annotation
learning. The image annotations in Profiset have no fixed
vocabulary and their quality is not centrally supervised. At the same time, however, the photographers are
interested in selling their photos and are thus motivated to provide rich sets of relevant keywords.

As the second collection, we employed the 500K set of training images provided
by organizers (the SCIA trainset). The Profiset represents a large collection of general-purpose images with as precise annotations
as can be achieved in a non-controlled environment. The SCIA trainset is smaller
and the quality of text data is much lower; on the other hand, it has been designed to contain images for all keywords from the SCIA task concept lists, which
makes it a very good fallback for topics not sufficiently covered in Profiset.

\subsubsection{Parameters}

Important factors that influence the performance of search-based annotation are the
reference collection size, reliability of reference image annotations, and the quality of visual similarity
measure. In the DISA-MU submissions to the ImageCLEF contest, the visual similarity of images was
measured by a weighted combination of five MPEG7
global visual descriptors, which compare the distribution of colors and edges in the image (detailed description
of the descriptors and a distance function we used can be found in~\cite{Lokoc12,MPEG7}. 

For each query image, a fixed
number $k$ of most similar images was selected from one or both of the datasets and used
for further processing. The number $k$ needed to be chosen carefully, as it influences
the quality of results. If we could suppose that all found objects are relevant for the
query, a high $k$ would be advantageous. However, this is often not the case in similarity-based image retrieval,
where semantically irrelevant images are likely to be evaluated as visually similar to the query.
It was therefore necessary to experimentally determine such $k$ that the selected images provided sufficient amount of information
but did not introduce too much noise.


\subsection{Phase 2: Semantic Analysis}

In the second phase of the annotation process, the descriptions of images returned by
content-based retrieval need to be analyzed and linked to SCIA concepts of a given query
to decide about their (ir)relevance. During this phase, our
solution relies mainly on the WordNet semantic structure~\cite{WordNet}.
The words associated with similar images are first transformed into synsets. Next, a fixed number
of the most frequent synsets and the connecting WordNet semantic links to used to
construct a graph of candidate synsets, over which the probabilities of relevance are
computed. Finally, a fixed number of SCIA concepts connected to top-ranking synsets
is produced as the annotation output.

\subsubsection{Parameters}
Three important parameters need to be set for the semantic analysis phase: the maximum number $s$ of synsets to 
be considered per each word, the number $n$ of synsets that enter the graph-building phase, and the type of semantic links 
that are utilized in the graph. Again, the optimal setting of these parameters needs to balance the amount of
information gained from various sources and the level of noise introduced by non-relevant links.

\subsection{Results Achieved at ImageCLEF 2014}

After fine-tuning the various annotation parameters on SCIA development data, the DISA-MU team submitted several
competition runs to the competition.
The results of the ImageCLEF 2014 SCIA Task are summarized in Table~\ref{tab:NewResults},
more details can be found in~\cite{BudikovaCLEF14,CLEF14AnnotationOverview}. Altogether, the DISA team
ranked fifth out of eleven participating teams. The parameters used by our best-performing run DISA 4
are summarized in Table~\ref{tab:Parameters}.

\section{The DeCAF Similarity Search Module}

One of the crucial features of the MUFIN annotation system is its modularity,
which enables us to freely combine different processing modules~\cite{BatkoBBZ13}.
During the development of the solution for the DISA competition, we were already
working on a new module for similarity searching that uses recently published
DeCAF features~\cite{Donahue2014} for measuring visual distance of images.
Based on a very successful image classifier that exploits convolutional
neural networks~\cite{KrizhevskySH12}, these features have been shown to
perform promisingly in various image processing tasks. Therefore, we decided
to try them for our similarity search module. To the best of our knowledge,
the DeCAF features have not been previously used for similarity-based retrieval over large data, which
only increased our motivation for experimenting with these features in context
of search-based image annotation.

\subsection{DeCAF$_7$ Features}

The recent popularity of neural networks for image processing was triggered by the
neural network classifier developed by Alex Krizhevsky for the 2012 ImageNet
challenge, which defeated other participants of the contest by a significant margin~\cite{KrizhevskySH12}.
This convolutional neural network was trained on 1000 categories and 1M correctly classified examples
with the purpose of identifying these 1000 categories. However, it was
soon observed that intermediate outputs of hidden layers of the neural network can be
used as a feature for evaluation of image similarity in general~\cite{Donahue2014,KrizhevskySH12}. Although the classifier was
trained for a specific set of 1000 concepts, the derived features have been shown
to perform well when used as a basis for classification tasks with several different
target concept sets~\cite{Donahue2014}.

In our implementation, we utilize the DeCAF$_7$ feature, which is produced by the last
hidden layer of the neural network classifier developed by Krizhevsky. The neural network
has not been re-trained in any way, in particular the SCIA development data has not been
used to adjust the network parameters. 


\subsection{DeCAF Similarity Search}

The DeCAF$_7$ representation of a single image
consists of a 4096-dimensional vector of real numbers and its extraction is a
rather heavy computational task~\cite{Donahue2014}. However, once the descriptors are
extracted from a dataset, they can be efficiently indexed and searched. Specifically,
we employ the PPP-Codes technique~\cite{DNovakZ14}, which enables us to search a collection of 20M
images in 1-2 seconds. To compute the distance
of two image features, we utilize the standard Euclidean distance.

As we demonstrate in the following section, replacing the MPEG-similarity search module
by the DeCAF similarity search had immediate effect on quality of annotation results,
which was rapidly increased. However, we could also observe that different parameter
settings were suitable with DeCAF search than those we determined for MPEG7-based annotation. 
Therefore, we also study the relationships
between these two types of descriptors, the type and size of the searched dataset,
and some other annotation parameters.

\section{Evaluation}

\subsection{DISA DeCAF at ImageCLEF 2014}

Although the DeCAF component has been completed after the SCIA competition deadline, the organizers
kindly agreed to evaluate a new submission on the complete test set for us (out of the contest). 
Table~\ref{tab:NewResults} presents the SCIA competition results with this new run, denoted as DISA DeCAF.
Using the algorithm described in~\cite{CLEF14AnnotationOverview}, we recomputed the overall ranking
of participants. With the DISA DeCAF run, the DISA team would now rank as second while outperforming 
the winner in most sample-based quality measures.
Unfortunately, we cannot present all performance measures since the data provided by
competition organizers didn't provide sufficient information to compute the concept-based metrics
for different subsets of the test collection. However, it is clear that the KDEVIR solution
still significantly outperforms ours in terms of concept-based MF.

Annotation parameter settings utilized for the DISA DeCAF submission are summarized in Table~\ref{tab:Parameters}.
It can be observed that several parameter values differ from the settings used in DISA competition runs with
MPEG7 similarity. In the following sections, we discuss the parameters in more detail.

\subsection{Similarity Search Performance in Different Conditions}

\begin{figure}[t]
\begin{center}
\includegraphics[width=\textwidth]{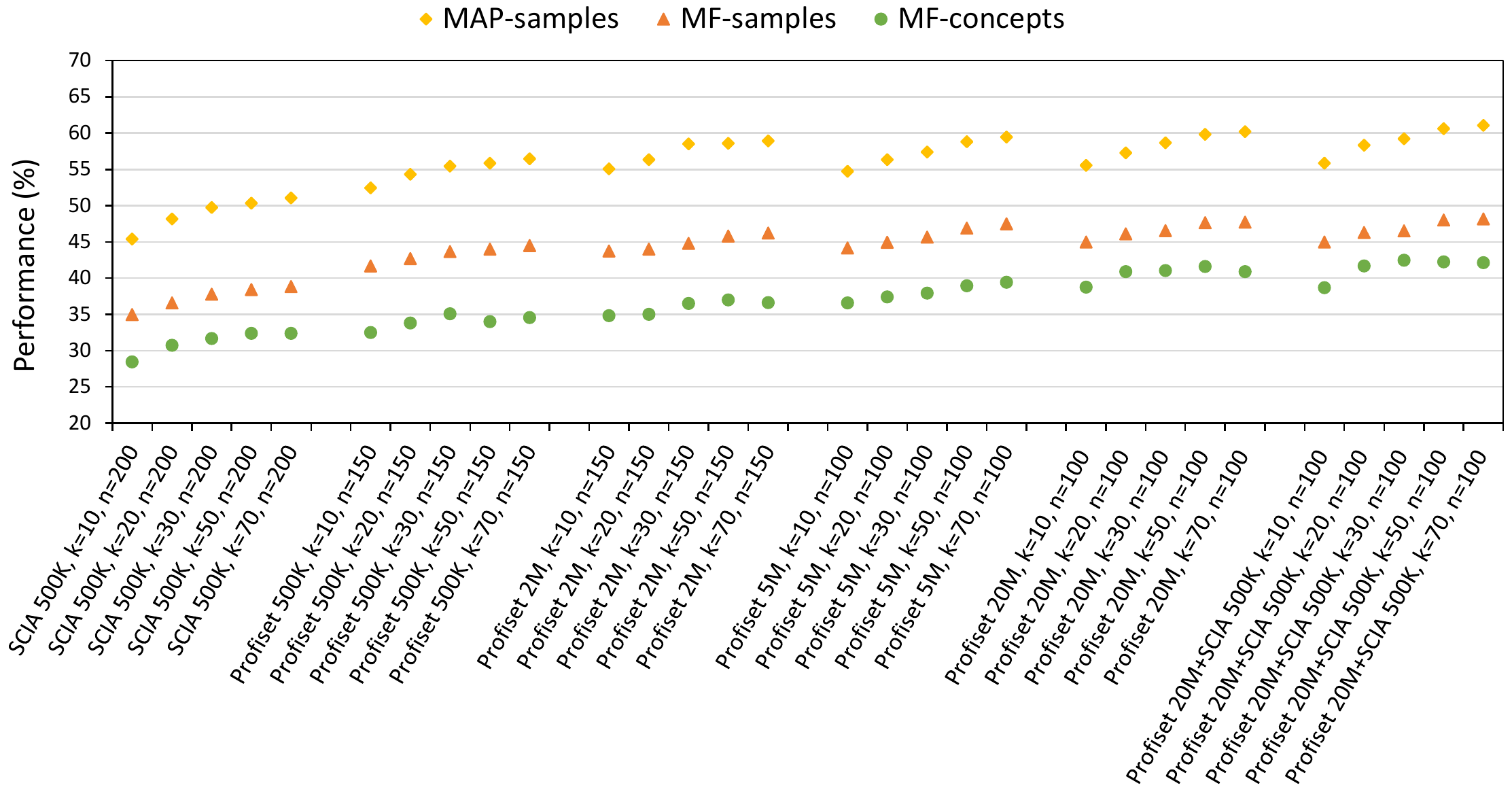}
\end{center}
\vskip -0.7\baselineskip
\caption{\label{fig:Graph} Influence of the dataset quality and size on the annotation performance. }
\end{figure}

To analyze the influence of dataset size and quality on the annotation system performance, we
utilized several test image collections that were employed in the similarity search phase. 
Apart from the SCIA 500K dataset and Profiset 20M, we created random subsets of Profiset
with 500K, 2M and 5M images. The performance of the annotation system on individual datasets
is depicted in Figure~\ref{fig:Graph}. For each set of experiments, optimal settings of the
semantic analysis phase were chosen so that the influence of similarity search parameters is
clearly visible.

The first two groups of results compare the performance of DeCAF on SCIA 500K and Profiset 500K.
We can clearly see that the higher-quality Profiset database provides better results in all three metrics. 
For both collections, the result quality grows with number $k$ of similar images taken
into consideration. 

The following result sets provide comparison of DeCAF performance on high-quality datasets of different sizes.
We can observe that increasing dataset size continually improves the result quality, so we can assume
that even better results could be achieved if we had a larger reference dataset with high-quality data. 
Again, better results are generally achieved for larger $k$.

Finally, the last group of results depicts the results achieved by combination of Profiset 20M and SCIA 500K data.
The slight improvement over Profiset 20M is caused by the fact that the SCIA 500K dataset covers all topics
considered in the annotation task. This increases the chance of correctly identifying less common concepts that do not
appear in the Profiset collection.

\renewcommand{\arraystretch}{1.2}
\renewcommand{\tabcolsep}{.11cm}
\begin{table}[t]\scriptsize
\caption{\label{tab:NewResults}The SCIA competition results table from~\cite{CLEF14AnnotationOverview} with a new line for DISA DeCAF results. Only
the best result for each group is given. The systems are ranked by overall performance as defined in~\cite{CLEF14AnnotationOverview}.}
\begin{tabular}{|l|c|c|c|c|c|c|c|c|c|c|c|c|c|}
\hline
\multirow{2}{*}{\parbox[t]{1.5cm}{System}} & \multicolumn{4}{c|}{MAP-samples} & \multicolumn{4}{c|}{MF-samples} & \multicolumn{5}{c|}{MF-concepts}\\ \cline{2-14}
 & all & ani. & food & 207 & all & ani. & food & 207 & all & ani. & food & 207 & unseen\\ \hline\hline
KDEVIR 9 & 36.8 & 33.1 & 67.1 &	28.9 & 37.7 & 29.9 & 64.9 &	32.0 & 54.7 & 67.1 & 65.1 &	31.6 & 66.1\\ \hline
\rowcolor{Gray}
DISA DeCAF & 48.6	& 51.0 & 67.2 &	32.3 & 39.9	& 44.4 & 48.5 & 26.7 & 41.1	& N/A & N/A	& N/A & 44.9\\ \hline
MIL 3 & 36.9 & 30.9 & 68.6 & 23.3 & 27.5 & 20.6	& 53.1 & 18.0 & 34.7 & 34.7 & 50.4 & 16.9 & 36.7\\ \hline
MindLab 1 & 37.0 & 43.1 & 63.0 & 22.1 & 25.8 & 17.0 & 45.2 & 18.3 & 30.7 & 35.1 & 35.3 & 16.7 & 34.7\\ \hline
MLIA 9 & 27.8 & 18.8 & 53.6 & 16.7 & 24.8 & 12.1 & 46.0 & 16.4 & 33.2 & 32.7 & 37.3 & 16.9 & 34.8\\ \hline
\rowcolor{Gray}
DISA 4 & 34.3 & 46.6 & 39.6 & 19.0 & 29.7 & 40.6 & 31.2	& 16.9 & 19.1 & 23.0 & 22.3 & 7.3 & 19.0\\ \hline
RUC 7 & 27.5 & 25.2 & 44.2 & 15.1 & 29.3 & 28.0 & 28.2 & 20.7 & 25.3 & 20.1 & 23.1 & 10.0 & 18.7\\ \hline
IPL 9 & 23.4 & 30.0 & 48.5 & 18.9 & 18.4 & 20.2 & 29.8 & 17.5 & 15.8 & 15.8 & 33.3 & 12.5 & 22.0\\ \hline
IMC 1 & 25.1 & 35.7 & 35.6 & 12.9 & 16.3 & 14.3 & 21.0 & 10.9 & 12.5 & 10.2 & 15.1 & 6.1 & 11.2\\ \hline
INAOE 5 & 9.6 & 6.9 & 15.0 & 8.5 & 5.3 & 0.4 & 0.5 & 6.4 & 10.3 & 1.0 & 0.8 & 17.9 & 19.0\\ \hline
NII 1 & 14.7 & 23.2 & 22.0 & 4.6 & 13.0 & 18.9 & 18.7 & 4.9 & 2.3 & 3.0 & 2.1 & 0.9 & 1.8\\ \hline
FINKI 1 & 6.9 & N/A & N/A & N/A & 7.2 & 8.1 & 12.3 & 4.1 & 4.7 & 6.3 & 9.0 & 2.9 & 4.7\\ \hline
\end{tabular}
\end{table}

\renewcommand{\arraystretch}{1.2}
\renewcommand{\tabcolsep}{.11cm}
\begin{table}[t]\scriptsize
\caption{\label{tab:Parameters}Annotation tool parameters in the competition run DISA-MU 04 and the new run DISA-MU DeCAF.}
\begin{tabular}{|m{2.9cm}|m{3.3cm}|C{2.5cm}|C{2.5cm}|}
\hline
\multirow{2}{*}{Annotation phase} & \multirow{2}{*}{Parameter} & \multicolumn{2}{c|}{Value} \\ \cline{3-4}
& & DISA-MU 04 & DISA-MU DeCAF \\ \hline \hline
\multirow{3}{*}{\parbox[t]{3.6cm}{Similar images retrieval}} & visual image descriptor & MPEG7 & DeCAF \\ \cline{2-4}
& datasets & \multicolumn{2}{c|}{both Profiset and SCIA trainset} \\ \cline{2-4}
& \# of similar images & 25 & 70\\ \hline
\multirow{3}{*}{Semantic analysis} & max \# of synsets per word & \multicolumn{2}{c|}{7} \\ \cline{2-4}
& \# of initial synsets	& 200 & 100 \\ \cline{2-4}
& relationships & \multicolumn{2}{c|}{hypernym, hyponym, holonym, meronym}\\ \hline
Final concepts selection & \# of best results & 7 & 5 \\ \hline
\end{tabular}
\end{table}

\subsection{Influence of Semantic Analysis}

Next, let us focus on the relationship between the performance of the first processing phase (the similarity search)
and the second phase (semantic analysis). Table~\ref{tab:Relationships} compares MPEG7-based and DeCAF-based similarity
search combined with different levels of semantic analysis. We can observe that the trends are consistent for both MPEG and DeCAF
-- in both cases, adding semantic analysis steps increases the final result quality. However, different parameters may be needed to achieve 
optimal results in combination with MPEG7 and DeCAF.

\renewcommand{\arraystretch}{1.2}
\renewcommand{\tabcolsep}{.11cm}
\begin{table}[t]\scriptsize
\caption{\label{tab:Relationships}Experiments on SCIA development dataset: MPEG and DeCAF similarity search over 20M Profiset combined with
different levels of semantic analysis.}
\begin{tabular}{|m{5.5cm}|c|c|c|c|c|c|c|}
\hline
Semantic analysis & MP-c &  MR-c &  MF-c &  MP-s &  MR-s & MF-s & MAP-s\\ \hline\hline
MPEG, basic frequency analysis & 18.2 & 32.9 & 19.0 & 23.8 & 40.8 & 27.6 & 34.7\\ \hline
MPEG, multiple meanings (synsets) per word, no relationships & 29.1 &  29.2  & 22.4 & 28.3 & 39.5 & 30.3 & 38.4\\ \hline
MPEG, multiple  meanings, hypernymy, hyponymy & 29.2 & 26.7 & 21.2 &  30.1 & 44.2 & 33.1 & 42.1\\ \hline
MPEG, multiple meanings, hypernymy, hyponymy, meronymy, holonymy & 29.5 & 27.5 & 21.8 & 30.4 & 45.2 & 33.5 & 42.7\\ \hline\hline
DeCAF, basic frequency analysis & 32.5 & 46.8 & 33.6 & 37.4 & 49.9 & 39.6 & 49.5\\ \hline
DeCAF, multiple meanings (synsets) per word, no relationships & 48.9 & 48.8 & 40.6 & 42.7 & 55.6 & 44.9 & 55.6\\ \hline
DeCAF, multiple meanings, hypernymy, hyponymy& 48.0 & 48.5 & 41.5 & 44.6 & 61.0 & 48.1 & 60.8\\ \hline
DeCAF, multiple meanings, hypernymy, hyponymy, meronymy, holonymy & 47.7 & 49.0 & 41.7 & 44.7 & 61.5 & 48.3 & 61.1 \\ \hline
\end{tabular}
\end{table}

The first parameter that influences the performance of the semantic analysis is the number $s$ of candidate synsets
considered for each word produced by the similarity search. This parameter behaves consistently for both MPEG and DeCAF
similarity, with the optimal value of $s$ being 7 in both cases. Similarly, adding WordNet semantic links to the
candidate synset graph improves the annotation results in combination with both similarity measures. 
However, the optimal number $n$ of synsets that should enter the graph-building phase differs for MPEG and DeCAF
and also for different datasets employed in the similarity search phase. 

With the 20M Profiset collection, DeCAF-based annotation performs best with $n=100$, whereas MPEG-based
annotation requires $n=200$. We can conclude that the DeCAF-based search produces images that are more semantically relevant to the query,
therefore the initial frequency-based ordering of synsets is already rather good.
Synsets with lower rank are less likely to be relevant to image topic and rather introduce
noise into the semantic processing. This observation is also important from the efficiency point of view --
less initial synsets form a smaller semantic graph, which implies faster execution of the semantic analysis phase. 
Moreover, the optimal value of $n$ also depends on the size and quality of the reference dataset. Specifically,
our experiments show that 200 synsets should be used for the SCIA 500K dataset, whereas the Profiset 500K and Profiset 2M
require 150 initial synsets.

\subsection{Efficiency}

Finally, let us briefly discuss the computation costs of the annotation process.
On average, each image takes about 4-5 seconds to process. The overall processing time
is determined by the costs of four computationally intensive phases: 1) extraction of DeCAF
features from the query image, 2) the similarity search, 3) retrieval of words for similar images
(these are not stored in the PPP-Codes index to minimize the index size), and 4) the
computation of synset probabilities over the candidate synset graph. The costs of individual
phases with the parameters utilized by DISA DeCAF submission are summarized in Table~\ref{tab:Costs}.

The current implementation offers near real-time response and can be further optimized in future.
In particular, the feature extraction can be made faster by introducing GPU processing,
while SSD disks can be used for keyword data storage. We will also focus on a more efficient 
implementation of the semantic analysis phase.

\renewcommand{\arraystretch}{1.2}
\renewcommand{\tabcolsep}{.11cm}
\begin{table}[t]\scriptsize
\centering
\caption{\label{tab:Costs}Computation costs for the DISA DeCAF annotation system.}
\begin{tabular}{|l|c|}
\hline
Phase & Time [s]\\ \hline\hline
Extraction of DeCAF features & 1\\ \hline
Similarity search in 20M images & 1-2\\ \hline
Retrieval of words for 70 most similar images & 1\\ \hline
Semantic analysis with 100 initial synsets & 0.5-1\\ \hline
\end{tabular}
\end{table}

%
%
%


\section{Conclusions and Future Work}%

In this paper, we have presented the results achieved by the DISA annotation system
with a DeCAF-based similarity search component. In comparison with our former system
that competed in the ImageCLEF Scalable Concept Annotation task, the DISA DeCAF 
quality of results is 10-20\,\% higher (depending on the metric). If the DISA DeCAF
submission was ready in time of the contest, DISA-MU would have placed second in the overall ranking
of participants.

The evaluation results also show that DISA DeCAF achieved better results than some other
groups who also employed the neural network approach. This confirms the importance of the semantic
analysis step developed by our group. Furthermore, we have demonstrated high adaptability of
our system, which can be easily adjusted to other application domains by simply replacing
the similarity evaluation function.

The SCIA overview paper~\cite{CLEF14AnnotationOverview} poses a question of whether the
overlap between ImageNet concepts (which were used for DeCAF definition) and SCIA concepts may bias the results of systems that
employ the DeCAF features. While it is true that some concepts appear both in ImageNet and SCIA
lists, we believe that this is not significant as any image descriptor is likely to be
trained on a similar set of common visual concepts. Nonetheless, future experiments
can be designed to test the DISA annotation system with DeCAF-like features derived
from a neural network trained on non-overlapping concepts.


 \vspace{-1ex}
 \section*{Acknowledgments}
 \vspace{-1ex}
{\small
This work was supported by the Czech national research project GBP103/12/G084.
The hardware infrastructure was provided by the METACentrum under the programme
LM 2010005.}

\newcommand{\BIBdecl}{\setlength{\itemsep}{11pt}}

\bibliographystyle{splncs03}
\bibliography{article}

\end{document}